\documentclass[onecolumn,showpacs]{revtex4}

\usepackage{graphicx}% Include figure files
\usepackage{dcolumn}% Align table columns on decimal point
\usepackage{bm}% bold math

\input epsf

\begin{document}

\title{\Large Collapse Dynamics of a Star of Dark Matter and Dark Energy}

\author{\bf~Subenoy~Chakraborty\footnote{subenoy@iucaa.ernet.in}~and
~Tanwi~Bandyopadhyay}

\affiliation{Department of Mathematics,~Jadavpur
University,~Calcutta-32, India.}

\date{\today}

\begin{abstract}
In this work, we study the collapse dynamics of an inhomogeneous
spherically symmetric star made of dark matter (DM) and dark
energy (DE). The dark matter is taken in the form of a dust cloud
while anisotropic fluid is chosen as the candidate for dark
energy. It is investigated how dark energy modifies the
collapsing process and is examined whether dark energy has any
effect on the Cosmic Censorship Conjecture. The collapsing star
is assumed to be of finite radius and the space time is divided
into three distinct regions $\Sigma$ and $V^{\pm}$, where
$\Sigma$ represents the boundary of the star and $V^{-}(V^{+})$
denotes the interior (exterior) of the star. The junction
conditions for matching $V^{\pm}$ over $\Sigma$ are specified.
Role of Dark energy in the formation of apparent horizon is
studied and central singularity is analyzed.
\end{abstract}


\maketitle

\section{\normalsize\bf{Introduction}}

The collapse dynamics in the framework of Einstein gravity was
started long ago by Oppenheimer and Snyder [1]. An extensive study
of gravitational collapse for matter cloud in the form of
pressure-less dust has been done in recent years [2-6] (for
recent reviews see [7]). These studies show the validity of
Cosmic Censorship Conjecture in six and higher dimensions under
reasonable physical conditions, the role of initial data on the
end state of collapse and some dynamical symmetries [8] on the
initial data set. Also the collapse dynamics has considerable
astrophysical significance.\\

So far, these studies are mostly concentrated to dust cloud and
matter with pressure components [9-13]. Also there are few works
[14-15] where in addition to dust matter, one has cosmological
constant with it. Recently, Madhav, Goswami and Joshi [16] have
studied in details the effect of a cosmological constant in the
background of asymptotically anti de-Sitter (or de-Sitter) space
-time. They have treated the cosmological constant as a dark
energy component, motivated by the recent astronomical
observations [17] of high redshift type Ia Supernova. Usually, the
dark energy behaves as the source of repulsive gravity and is
considered to be important in the present era to justify the
undergoing accelerated expansion of the universe.\\

Thus, it is interesting to study gravitational collapse of a
matter cloud containing dark energy. In the present work, we
examine the collapsing process of an~inhomogeneous spherically
symmetric star whose matter inside contains a combination of dark
matter (dust) and dark energy. We have studied in two different
sections the collapsing scenario for~TBL model and general
spherically symmetric model. The formation of apparent horizon
has been investigated in details for various cases and it is
examined whether bouncing solutions are possible or not. The
paper ends with a short discussion and concluding remarks.\\

\section{\normalsize\bf{Basic Equations}}

The metric~ansatz of a general spherically symmetric metric can
be written as

\begin{equation}
ds^{2}=-e^{2\gamma}dt^{2}+e^{2\alpha}dr^{2}+R^{2}d\Omega^{2}
\end{equation}

where $\alpha, \gamma$ and $R$ are functions of $t$ and $r$ and
$d\Omega^{2}=d\theta^{2}+\sin^{2}\theta d\phi^{2}$ is the usual
line element on a unit $2$-sphere. The form of the energy momentum
tensor is taken as

\begin{equation}
{T_{\mu}^{\nu}}_{(total)}={T_{\mu}^{\nu}}_{(DM)}+{T_{\mu}^{\nu}}_{(DE)}
\end{equation}

Here the coordinates ($t,r,\theta,\phi$) are taken to be~comoving
and the components of the above energy momentum tensor in this
comoving frame are

\begin{equation}
{T_{\mu}^{\nu}}_{(DM)}=diag(-\rho_{_{DM}},0,0,0)
\end{equation}

and

\begin{equation}
{T_{\mu}^{\nu}}_{(DE)}=diag(-\rho,p_{r},p_{t},p_{t})
\end{equation}

The quantities $\rho_{_{DM}},\rho,p_{r}$ and $p_{t}$ are
respectively the dark matter density, dark energy matter density,
radial and tangential pressure for dark energy and are functions
of $r,t$. In order to satisfy the weak energy condition for the
whole matter we have

\begin{equation}
{T_{\mu}^{\nu}}_{(total)}V^{\mu}V^{\nu}\geq0
\end{equation}

for any time-like vector $V^{\mu}$. In explicit form this means

\begin{equation}
\rho_{_{T}}\geq0,~~\rho_{_{T}}+p_{r}\geq0,~~\rho_{_{T}}+p_{t}\geq0
\end{equation}

where $\rho_{_{T}}=\rho_{_{DM}}+\rho$, is the total matter
density. Now the explicit form of the Einstein's field equations
for the metric (1) with matter field given by (2) are (choosing
$8\pi G=c=1$)

\begin{equation}
e^{-2\gamma}\left(\frac{\dot{R}^{2}}{R^{2}}+2\dot{\alpha}\frac{\dot{R}}{R}\right)
+\frac{1}{R^{2}}-e^{-2\alpha}\left(2\frac{R''}{R}+\frac{R'^{2}}{R^{2}}-2\alpha'\frac{R'}{R}\right)
=\rho_{_{T}}
\end{equation}

\begin{equation}
-e^{-2\gamma}\left(2\frac{\ddot{R}}{R}+\frac{\dot{R}^{2}}{R^{2}}-2\dot{\gamma}\frac{\dot{R}}{R}\right)
-\frac{1}{R^{2}}+e^{-2\alpha}\left(\frac{R'^{2}}{R^{2}}+2\gamma'\frac{R'}{R}\right)=p_{r}
\end{equation}

\begin{equation}
-e^{-2\gamma}\left(\ddot{\alpha}+\dot{\alpha}^{2}-\dot{\alpha}\dot{\gamma}+\frac{\ddot{R}}{R}
+\dot{\alpha}\frac{\dot{R}}{R}-\dot{\gamma}\frac{\dot{R}}{R}\right)+e^{-2\alpha}\left(\gamma''
+{\gamma'}^{2}-\alpha'\gamma'+\frac{R''}{R}+\gamma'\frac{R'}{R}-\alpha'\frac{R'}{R}\right)=p_{t}
\end{equation}

and

\begin{equation}
\dot{R}'-\dot{\alpha}R'-\gamma'\dot{R}=0
\end{equation}

where over dot(.) and dash(') stand for partial derivatives with
respect to $t$ and $r$ respectively. We now introduce the mass
function $m(t,r)$, defined by

\begin{equation}
m(t,r)=R(1+e^{-2\gamma}\dot{R}^{2}-e^{-2\alpha}{R'}^{2})
\end{equation}

which can be interpreted as the total mass inside the~comoving
radius $r$ at the instant $t$. This definition was first
introduced by Cahill and McVittie [18] and since then it has been
widely used in different context. Also, for asymptotically flat
space times, it corresponds to the correct Bondi mass at
infinity. If the boundary of the collapsing cloud is leveled by
the~comoving coordinate $r_{\Sigma}$, then the total mass of the
collapsing star at the instant $t$ is given by

\begin{equation}
m_{\Sigma}(t)=m(t,r_{\Sigma})
\end{equation}

For regularity on the initial hypersurface $t=t_{i}$, we must
choose $m(t_{i},0)=0$.\\

Using this mass term, the field equations (7) and (8) can be
written in compact form as

\begin{equation}
\rho_{_{T}}=\frac{m'}{R^{2}R'},~~~~~~~~~~~~~~p_{r}=-\frac{\dot{m}}{R^{2}\dot{R}}
\end{equation}

As the matter field is a mixture of dark matter (inhomogeneous
dust) and dark energy (anisotropic fluid), so from energy momentum
conservation relation ${{{T_{\mu}}^{\nu}}_{;}}_{\nu}=0$, we obtain

\begin{equation}
\dot{\rho_{_{T}}}+\rho_{_{T}}\left(\dot{\alpha}+2\frac{\dot{R}}{R}\right)
+\dot{\alpha}p_{r}+2p_{t}\frac{\dot{R}}{R}=0
\end{equation}

and

\begin{equation}
p_{r}'+\left(\rho_{_{T}}+p_{r})\gamma'-2(p_{t}-p_{r}\right)\frac{R'}{R}=0
\end{equation}

If $Q(t,r)$ denotes the interaction between the dust cloud and
dark energy, then a self consistent conservation equation for
each of them can be separately written as (see eqn(14))

\begin{equation}
\dot{\rho_{_{DM}}}+\left(\dot{\alpha}+2\frac{\dot{R}}{R}\right)\rho_{_{DM}}=Q(t,r)
\end{equation}

and

\begin{equation}
\dot{\rho}+\left(\dot{\alpha}+2\frac{\dot{R}}{R}\right)\rho+\dot{\alpha}p_{r}
+2\frac{\dot{R}}{R}p_{t}=-Q(t,r)
\end{equation}

Thus we have six independent differential equations namely
equations (10), (11), (13), (16) and (17) containing eight
unknowns (geometrical or physical) namely, $\rho_{_{DM}}, \rho,
p_{r}, p_{t}, Q(t,r), \alpha, \gamma$ and $R$, giving us freedom
to choose two free functions. In particular, for a given initial
data satisfying the weak energy condition, the choice of these
free functions will determine the matter distribution and metric
of the space time and hence a particular dynamical evolution of
the initial data.\\

On the initial hypersurface: $t=t_{i}$ we have to specify eight
functions of $r$ namely $\rho_{_{DM_{0}}}(r), \rho_{0}(r),
p_{r_{0}}(r), p_{t_{0}}(r), \gamma_{0}(r), \alpha_{0}(r),
R_{0}(r)$ and $Q_{0}(r)$. By proper scaling of the radial
coordinate $r$, we can choose $R_{0}(r)=r$, without any loss of
generality. It is to be noted that the remaining seven (initial
data) functions are not independent but are related by the field
equations. Further, to preserve regularity and smoothness of the
initial data, we have to assume that the initial pressures at the
regular centre $r=0$ should satisfy\\

i) gradient of the initial pressures vanish at the centre i.e, $p_{r_{0}}'(0)=p_{t_{0}}'(0)=0$\\

ii) the initial pressure should be isotropic at the centre i.e,
$p_{r_{0}}(0)=p_{t_{0}}(0)$.\\

From the field equations (13), we see that the total matter
density becomes infinity either when $R=0$ or $R'=0$ or both. The
case for $R'=0$ is termed as shell crossing singularity which is
not of much physical interest as it is gravitationally weak in
nature and is removable singularity. The other one (which we
shall discuss here) is termed as shell focusing singularity where
the physical radius of all matter shells goes to zero ($R=0$). So
we shall assume $R'>0$ at all instant to avoid any crossing of
the shells. If $t_{s}(r)$ denotes the instant when a shell at
coordinate radius $r$ collapse to $r=0$, then we have
$R(t_{s}(r),r)=0$. ( Note that the singularity time $t_{s}$ is a
function of $r$ due to the~inhomogeneity of the space-time ). In
the subsequent sections we shall consider the collapse dynamics
for two separate assumptions namely\\

i) a geometrical assumption: $\gamma=0$ (LTB model) and\\

ii) a physical assumption: $p_{r}=0$.\\

\section{\normalsize\bf{Collapsing process in LTB model with DM and
DE}}

The metric of the space time describing the collapsing matter in
the standard Lemaitre-Tolman-Bondi (LTB) model is given by

\begin{equation}
ds^{2}=-dt^{2}+\frac{R'^{2}}{1+f(r)}dr^{2}+R^{2}d\Omega^{2}
\end{equation}

The simplified field equations are

\begin{equation}
\frac{\dot{R}^{2}}{R^{2}}+2\frac{\dot{R}}{R}\frac{\dot{R}'}{R'}-\frac{f}{R^{2}}
-\frac{f'}{RR'}=\rho_{_{T}}
\end{equation}

\begin{equation}
\frac{f}{R^{2}}-\frac{\dot{R}^{2}}{R^{2}}-2\frac{\ddot{R}}{R}=p_{r}
\end{equation}

and

\begin{equation}
\frac{f'}{2RR'}-\frac{\dot{R}}{R}\frac{\dot{R}'}{R'}-\frac{\ddot{R}}{R}
-\frac{\ddot{R}'}{R'}=p_{t}
\end{equation}

with energy conservation equations

\begin{equation}
\dot{\rho_{_{DM}}}+\left(2\frac{\dot{R}}{R}+\frac{\dot{R}'}{R'}\right)\rho_{_{DM}}=Q(t,r)
\end{equation}

\begin{equation}
\dot{\rho}+\left(2\frac{\dot{R}}{R}+\frac{\dot{R}'}{R'}\right)\rho+p_{r}\frac{\dot{R}'}{R'}
+2p_{t}\frac{\dot{R}}{R}=-Q(t,r)
\end{equation}

and

\begin{equation}
p_{r}'=2(p_{t}-p_{r})\frac{R'}{R}
\end{equation}

From equation (24), it is to be noted that\\

a) anisotropy of pressure implies the inhomogeneity of the
pressure and vice-versa\\
b) dark energy model with vanishing radial pressure and non-zero
tangential pressure ($p_{r}=0, p_{t}\neq0$) is not possible.\\

From the recent study of the role of pressure in quasi-spherical
gravitational collapse, we choose the radial pressure as [12]

\begin{equation}
p_{r}(t,r)=-\frac{g(r)}{R^{n}}
\end{equation}

Then from the conservation equation (24), the tangential pressure
has the expression

\begin{equation}
p_{t}(t,r)=-\frac{g(r)}{R^{n}}\left(1-\frac{n}{2}\right)-\frac{g'(r)}{2R^{n-1}R'}
\end{equation}

(Note that we have taken $-$ve sign for $p_{r}$ as pressure is
$-$ve for dark energy). The arbitrary function $g(r)$ should be
$\bigcirc(r^{n+2})$ near $r=0$ for the regularity and smoothness
of the initial pressure. Solving the field equations (13) we have
the expression for mass function and total matter density as

\begin{equation}
m(t,r)=m_{0}(r)+\frac{g(r)}{(3-n)R^{n-3}},~~~~~~~~~~(n\neq3)
\end{equation}

and

\begin{equation}
\rho_{_{T}}(t,r)=\frac{m_{0}'(r)}{R^{2}R'}+\frac{g'(r)}{(3-n)R^{n-1}R'}+\frac{g(r)}{R^{n}}
\end{equation}

where $m_{0}(r)$ is an arbitrary function of $r$, restricted by
the energy conditions. The evolution equation for the area radius
$R$ can be obtained from the field equation (20) (or from the
definition of the mass function (11)) as

\begin{equation}
\dot{R}^{2}=f(r)+\frac{m_{0}(r)}{R}+\frac{g(r)}{(3-n)R^{n-2}}
\end{equation}

Now using the scaling independence of the coordinate $r$ [16] let
us write

\begin{equation}
R(t,r)=rv(t,r)
\end{equation}

so that

\begin{equation}
v(t_{i},r)=1,~~~~~~~~~~v(t_{s},r)=0,~~~~~~~~~\dot{v}<0
\end{equation}

Then from (29) the time evolution equation for $v$ is

\begin{equation}
\dot{v}^{2}=\left[\frac{f(r)}{r^{2}}v+\frac{m_{0}(r)}{r^{3}}
+\frac{g(r)}{(3-n)r^{n-1}}v^{3-n}\right]/v=\frac{V(r,v)}{v}
\end{equation}

This is the evolution equation of a particular shell and $V(r,v)$
may be termed as an effective potential.\\

One of the main questions that we shall address here is how the
collapse dynamics gets modified due to the presence of dark
energy. In case of pure dust collapse, once the collapse
initiates, starting from an initial regular hypersurface, the
cloud necessarily collapses to a singularity due to gravitational
attraction without any reversal or bounce. On the other hand, the
negative pressure of the dark energy produces repulsive
gravitational force and slows down the collapsing process. So it
is expected that, due to the presence of dark energy, the
collapsing process may not be smooth. Thus, it is interesting to
study in details the qualitative nature of the effective
potential $V(r,v)$ for which the allowed regions of motion
correspond to $V(r,v)\geq0$. In fact initially, when the collapse
starts, we have $\dot{v}<0$ and we will have a rebounce if we get
$\dot{v}=0$ before the shell has become singular. In other words,
study of various evolution for a particular shell is equivalent to
study the nature of the roots of the equation $V(r,v)=0$ for
fixed $r$.\\

To get a clear idea, let us consider a smooth initial data for
which initial density, radial pressure and curvature are smooth
near $r=0$ and the corresponding functions will have power series
expansion as follows:

\begin{equation}
\left.
\begin{array}{ll}
\rho_{_{T0}}(t_{i},r)=\rho_{00}+\rho_{02}r^{2}+\rho_{04}r^{4}+........\\\\
p_{r0}(t_{i},r)=-{p_{02}}^{(r)}r^{2}-{p_{04}}^{(r)}r^{4}-{p_{06}}^{(r)}r^{6}-
.........\\\\
f(r)=f_{2}r^{2}+f_{4}r^{4}+........
\end{array}
\right\}
\end{equation}

As we are concentrating on the evolution of the shells near $r=0$
so we may neglect here higher order terms in the above expansions.
Thus near the centre ($r<<r_{\Sigma}$) equation (32) becomes

\begin{equation}
\dot{v}^{2}=\left[f_{0}v+\frac{\rho_{00}}{3}+\frac{{p_{02}}^{(r)}r^{3}}{(3-n)}v^{3-n}\right]/v
\end{equation}

The numerator of the right hand side of the above equation (i.e
the effective potential $V(r,v)$) is a polynomial in $v$ and in
general may or may not have positive real roots which correspond
to physical cases. As $V(r,0)=\frac{\rho_{00}}{3}>0$ so any
region between $R=rv=0$ and the first positive zero of $V(r,v)$
always becomes singular during collapse while if there are two
consecutive positive real roots then the region between them is
forbidden as $\dot{v}^{2}<0$ there. Hence a particular shell will
bounce if it lies in the initial epoch ($v=1$) to the right of
the second positive root. We shall present various possibilities
for $n=1$ and derive the necessary conditions below:\\

{\bf a)$f_{0}=0$:~(Marginally Bound Case)}\\

Here as both $\rho_{00}$ and ${p_{02}}^{(r)}$ are positive so
both roots of the quadratic equation are complex and hence
$\dot{v}^{2}$ is always positive and no bounce occurs. Hence a
singularity always forms from the initial collapse.\\

{\bf b)$f_{0}>0$}:\\

Both roots will be real and negative if
${f_{0}}^{2}\geq\frac{2}{3}\rho_{00}~{p_{02}}^{(r)}r^{2}$,
otherwise they are complex in nature.\\

{\bf c)$f_{0}<0$}:\\

For the restriction
${f_{0}}^{2}\geq\frac{2}{3}\rho_{00}~{p_{02}}^{(r)}r^{2}$, both
roots are real and positive, otherwise they are complex. So
bounce will be possible in this case.\\

{\bf Horizons}:\\

For observers at infinity, the event horizon plays an important
role in characterizing the nature of the singularity. As the
calculations are complex in nature, so instead of event horizon,
one considers a trapped surface which is a compact space-time
2-surface with normals on both sides as future pointing
converging null geodesic families. In fact, if ($r=$constant,
$t=$constant) the 2-surface $S_{r,t}$ is a trapped surface then
it and its entire future development lie behind the event horizon
provided the density falls off fast enough at infinity. As
apparent horizons (AH) are the boundaries of trapped regions
(surfaces), so in the present case it can be written as [19]\\

~~~~~~~~~~~~~~~~AH : ~~~~~~~~~$g^{\mu\nu}R_{,\mu}R_{,\nu}=0$\\

Using the metric form (1) one gets\\

~~~~~~~~~~~~~~$-e^{-2\gamma}\left(\frac{\partial R}{\partial
t}\right)^{2}+e^{-2\alpha}\left(\frac{\partial R}{\partial r}\right)^{2}=0$,\\

which with the help of the mass function simplifies to

\begin{equation}
1-\frac{m(t,r)}{R}=0
\end{equation}

In the present case (i.e, for LTB model ) the apparent horizon is
characterized by

\begin{equation}
1-\frac{m_{0}(r)}{R}-\frac{g(r)}{3-n}\frac{1}{R^{n-2}}=0~~~~~~~~~~~(n\neq3)
\end{equation}

We shall now discuss the formation of apparent horizon for
different values of $n$:\\

{\bf $n=1$}:\\

In this case, equation (36) is a quadratic equation in $R$,
having real positive roots if \\

(i)~~~$0<m_{0}g<\frac{1}{2}$~~~ and~~~ (ii)~~~$g(r)>0$\\

and these horizons are given by

\begin{equation}
\left.
\begin{array}{ll}
R_{c}(r)=\left\{1-\sqrt{1-2m_{0}g}\right\}/g\\\\
R_{b}(r)=\left\{1+\sqrt{1-2m_{0}g}\right\}/g
\end{array}
\right\}
\end{equation}

These are usually termed as the cosmological and the black hole
horizons [19].\\

{\bf $n=2$}:\\

Equation (36) reduces to a linear equation in $R$, so there is
only one horizon namely\\

~~~~~~~~~~~~~~~~$R_{bc}(r)=\frac{m_{0}(r)}{1-g(r)}$~~~~~~~~~~
(assuming $g(r)<1$).\\

{\bf $n=4$}:\\

Again equation (36) is a quadratic equation in $R$, having two
horizons if $|m_{0}|>2\sqrt{g}$.\\

{\bf $n=5$}:\\

Here equation (36) simplifies to a cubic equation in $R$ as

\begin{equation}
2R^{3}-2m_{0}(r)R^{2}+g(r)=0
\end{equation}

In the following we shall discuss the nature of the roots of the
cubic equation for various possibilities:\\

i) For
$g(r)<\frac{4{m_{0}}^{3}}{27}\left(1+\frac{1}{3\sqrt{3}}\right)$,\\

there are two positive roots of (38), which correspond to two
apparent horizons namely the cosmological horizon and the black
hole horizon given by

\begin{equation}
\left.
\begin{array}{ll}
R_{c}(r)=\frac{m_{0}(r)}{3}+\frac{2{m_{0}}^{3}}{(27)^{3/2}}
Cos\left[\frac{1}{3}Cos^{-1}\left\{-\frac{(27)^{3/2}}{2{m_{0}}^{3}}
\left(\frac{g}{2}-\frac{2{m_{0}}^{3}}{27}\right)\right\}\right]\\\\
R_{b}(r)=\frac{m_{0}(r)}{3}+\frac{2{m_{0}}^{3}}{(27)^{3/2}}
Cos\left[\frac{4\pi}{3}+\frac{1}{3}Cos^{-1}\left\{-\frac{(27)^{3/2}}{2{m_{0}}^{3}}
\left(\frac{g}{2}-\frac{2{m_{0}}^{3}}{27}\right)\right\}\right]
\end{array}
\right\}
\end{equation}

ii) When
$g(r)=\frac{4{m_{0}}^{3}}{27}\left(1+\frac{1}{3\sqrt{3}}\right)$,\\

then there is only one positive root (corresponds to a single
apparent horizon) of the above cubic equation given by

\begin{equation}
R_{bc}(r)=\frac{m_{0}}{3}+{m_{0}}^{3}/(27)^\frac{3}{2}
\end{equation}

iii) If $g(r)>\frac{4{m_{0}}^{3}}{27}\left(1+\frac{1}{3\sqrt{3}}\right)$,\\

then there are no possible roots and hence there are no apparent
horizons.\\

Further, if $t_{ah}(r)$ be the time of formation of apparent
horizon then from the evolution equation (34) we have

\begin{equation}
t_{ah}(r)-t_{s}(r)=-{\int_{0}}^{v_{_{ah}}}\frac{\sqrt{v}dv}{\left[f_{0}v
+\frac{\rho_{_{00}}}{3}+\frac{{p_{_{02}}}^{(r)}r^{3}}{3-n}v^{3-n}\right]^{\frac{1}{2}}}
\end{equation}

where we have considered shells close to $r=0$. Here $t_{s}(r)$
is the time of formation of singularity of a shell at comoving
radius $r$ and $v_{ah}=R_{b}$ (or $R_{c})/r$.\\

It is to be noted that the presence of the dark energy in the
form of anisotropic fluid characterizes the formation of apparent
horizon and also influences the time difference between the
formation of apparent horizon and singularity formation.\\

The above time difference will characterize the final end state
of gravitational collapse and also the nature of the resulting
singularity. If the formation of the horizon precedes the
formation of the central singularity, then the singularity will
necessarily be covered, i.e, black hole will be formed. On the
other hand, if the time difference is reversed then end state of
collapse leads to a naked singularity. As this characterization
of the singularity is local, so for global visibility, we should
examine whether it is possible to have future directed null
geodesics that terminate at the singularity in the past. Since
analysis of null geodesic has been done widely so we are not presenting
it here.\\

\section{\normalsize\bf{General spherically symmetric collapsing model
with} $p_{r}=0, p_{t}\neq0$ \bf{for dark energy}}

Over the last few years spherically symmetric collapse with
anisotropic pressure in the form of vanishing radial pressure
have been studied in details. The main objective of the present
study is to examine whether bouncing situation is possible due to
the presence of dark energy and to investigate singularity
formation conditions. Also we try to understand how dark energy
affects the junction conditions at the boundary of the collapsing
star to an exterior region, the formation of trapped surfaces
(apparent horizon) and the nature of the central shell focusing
singularity.\\

For general spherically symmetric model (given by equation (1)),
if we assume the radial pressure to be zero then from the field
equations (13) we have

\begin{equation}
\rho_{_{T}}(r,t)=\frac{m'(r)}{R^{2}R'}
\end{equation}

and from the conservation equation (15) using (42) we get

\begin{equation}
p_{t}=\frac{m'(r)\gamma'}{2RR'^{2}}
\end{equation}

Here $m(r)$ is an arbitrary function of $r$, satisfying the
energy conditions. For the remaining freedom to choose one
function, we take the metric coefficient $\gamma(r,t)$ in the
specific form

\begin{equation}
\gamma(r,t)=\gamma_{0}(t)+\beta(R)
\end{equation}

(motivation for choosing such a form is given in ref. [16]).\\

Using this choice of $\gamma$ in equation (10) we have the other
metric coefficient as

\begin{equation}
e^{\alpha}=\frac{R'(t,r)}{b(r)e^{\beta(R)}}
\end{equation}

with $b(r)$ an arbitrary $C^{2}$ function of $r$.\\

These choices of the metric coefficients when substituted in the
definition of the mass function (i.e, equation (11)), we obtain
the evolution equation for the area radius $R$ as

\begin{equation}
\dot{R}^{2}=\frac{e^{2\beta(R)}}{R}\left[m(r)+\left\{b^{2}(r)e^{2\beta(R)}-1\right\}R\right]
\end{equation}

where a multiplicative function of time has been eliminated by a
suitable scaling of the time coordinate. As in the previous
section, introducing $R(t,r)=rv(t,r)$ the above equation
transforms to

\begin{equation}
\dot{v}^{2}=\frac{e^{2\beta(rv)}}{v}\left[\frac{m(r)}{r^{3}}
+\frac{\left\{b^{2}(r)e^{2\beta(rv)}-1\right\}}{r^{2}}v\right]
\end{equation}

To study this evolution equation near the singularity ($r=0$),
one should consider smooth initial data (i.e, initial density and
pressure) with power series expansion (near $r=0$)

\begin{equation}
\left.
\begin{array}{ll}
\rho_{i}=\rho_{_{T}}(t_{i},r)=\rho_{0}+\rho_{2}r^{2}+\rho_{4}r^{4}+.....\\\\
p_{ti}=p_{t}(t_{i},r)=-p_{t2}r^{2}-p_{t4}r^{4}-p_{t6}r^{6}-.....~~~~~~~\text{and}\\\\
b^{2}(r)=1+b_{02}(r)r^{2}+b_{04}(r)r^{4}+.....
\end{array}
\right\}
\end{equation}

where the first term in the series expansion for $b^{2}(r)$ is
chosen to be unity as the metric looks Minkowskian near the
centre $r=0$. Thus close to $r=0$, neglecting higher order terms
in the above expression (47) for $\dot{v}$, one gets

\begin{equation}
\dot{v}^{2}=\left(1-\frac{2p_{t2}}{\rho_{0}}r^{2}v^{2}\right)\left(\frac{\rho_{0}}{3}
+b_{02}v-\frac{2p_{t2}b(r)}{\rho_{0}}v^{3}\right)/v
\end{equation}

During the collapsing process the scale factor $v$ decreases from
unity (at the initial epoch) to zero (at the time of
singularity), so the first factor in the numerator of the right
hand side of (49) is initially positive (as it represents
$e^{2\gamma}$). Hence this factor is positive definite during the
collapse dynamics. Thus, whether a shell bounces or not is
completely determined by the second factor in the effective
potential.\\

The second factor, which is a cubic equation in $v$, has in
general three distinct roots. As positive real roots are only
physically admissible, so for different situations we examine
below whether positive roots are possible or not. As the term in
quadratic power of $v$ (i.e, $v^{2}$) is absent, so whenever all
the three roots are real, there should be at least one positive
and one negative root.\\

\newpage

\begin{center}
{\bf \text TABLE (Analysis of the roots of the equation)}
\\\vspace{.2cm}
\begin{tabular}{|c|c|c|} \hline\hline
~~~&~~~&~~~\\
~{\bf Condition}~~~&~~~{\bf Nature of the roots}~~~&~~~{\bf
Physical Consequences}\\
~~~&~~~&~~~\\
\hline\hline ~~~&~~~&~~~\\

(i)~$b_{02}(r)\geq0$~~~&~~~Exactly one possible root\\
 ~~~&~~~(say $\alpha(r)$) and the two\\
 ~~~&~~~other roots are either negative\\
 ~~~&~~~or complex conjugate~~~&~~~
        The physically allowed range of $v$ is\\
 ~~~&~~~&~~~$[0,\alpha]$. If $\alpha\geq 1$,\\
 ~~~&~~~&~~~then there will always be a\\
 ~~~&~~~&~~~singularity while $\alpha<1$ implies\\
 ~~~&~~~&~~~an unphysical situation initially as\\
 ~~~&~~~&~~~$\dot{v}^{2}<0$, i.e, all shells in\\
 ~~~&~~~&~~~the allowed dynamical space [$0,\alpha$]\\
 ~~~&~~~&~~~becomes singular starting from initial collapse\\
~~~&~~~&~~~\\
\hline ~~~&~~~&~~~\\

(ii)~$b_{02}(r)\leq0$\\
(a)~$\rho_{0}>\frac{2}{3}\frac{{b_{02}}^{3}(r)}{p_{t2}b(r)}$~~~&~~~
No positive real root~~~&~~~Singularity is always the final\\
 ~~~&~~~&~~~outcome of the collapse\\
~~~&~~~&~~~\\
\hline ~~~&~~~&~~~\\

(b)~$\rho_{0}<\frac{2}{3}\frac{{b_{02}}^{3}(r)}{p_{t2}b(r)}$~~~&~~~
Two positive roots $v_{1}(r)$\\
 ~~~&~~~and $v_{2}(r)$ ($v_{1}<v_{2}$) where\\
 ~~~&~~~$v_{1}=\lambda~Cos\psi$,\\
 ~~~&~~~$v_{2}=\lambda~Cos(\psi+\frac{4\pi}{3})$\\
 ~~~&~~~with
 $\lambda=\sqrt{\frac{2}{3}\frac{b_{02}}{p_{t2}b(r)}}$\\
 ~~~&~~~and
 $3\psi=Cos^{-1}\left[-\sqrt{\frac{3\rho_{_{0}}p_{t2}b(r)}{2{b_{02}}^{3}(r)}}\right]$
~~~&~~~Allowed space for dynamical evolution\\
~~~&~~~&~~~is when $v$ lies in the range $[0,v_{1}]$\\
~~~&~~~&~~~and $[v_{2},\infty)$. The value of $v$ in\\
~~~&~~~&~~~($v_{1},v_{2}$) is not allowed as\\
~~~&~~~&~~~$\dot{v}^{2}<0$. Initially, shells for \\
~~~&~~~&~~~$v\in[0,v_{1}]$ always become singular\\
~~~&~~~&~~~but shells for $v\in[v_{2},\infty)$ will undergo\\
~~~&~~~&~~~a bounce and subsequent expansion starts\\
~~~&~~~&~~~from initial collapse\\
~~~&~~~&~~~\\
\hline ~~~&~~~&~~~\\

(c)~$\rho_{0}=\frac{2}{3}\frac{{b_{02}}^{3}(r)}{p_{t2}b(r)}$~~~&~~~
Two equal positive roots~~~&~~~There is no forbidden region\\
~~~&~~~&~~~and bounce will occur if $\lambda<2$\\
~~~&~~~&~~~\\
\hline

\end{tabular}
\end{center}

\vspace{.2cm}

In case (b)(of (ii)), there are two positive real roots and
bounce will occur when the area radius approaches $R_{b}=rv_{2}$.
As in this case both singularity and bounce may occur so the
explicit conditions for a particular shell to become singular or
undergo a bounce can be stated as\\

i) For singularity: $\lambda~Cos\psi>1$\\

ii)For bounce: $\lambda~Cos(\psi+\frac{4\pi}{3})<1$\\

So far, we have analyzed the evolution of shells close to the
centre. But it is difficult to study (analytically) shells far
from the centre and simple expression for the singularity or
bounce is not possible, only one has to use numerical methods.
However, the above analysis can be applicable to the entire cloud
within the boundary $r=r_{\Sigma}$, provided either the initial
area radius $r_{\Sigma}$ is itself small compared to the initial
data coefficients, i.e, $\rho_{i}{r_{\Sigma}}^{i}$,
$p_{ti}{r_{\Sigma}}^{i}<<1$ or the initial data are such that
higher order coefficients in the power series expansion are
identically zero i.e, $\rho_{i}=0=p_{ti}, \forall~~ i\geq4$ and
$\rho_{2}{r_{\Sigma}}^{2}$, $p_{t2}{r_{\Sigma}}^{2}<<1$.
Moreover, if the whole collapsing cloud undergoes bounce starting
from the initial collapse (i.e, case(b) of (ii) holds for all $r$
in $0\leq r<r_{\Sigma}$) then the sufficient condition for
avoiding shell crossing is
$\frac{v_{2}(r+\delta)}{v_{2}(r)}\geq1, \forall~~
r~\in~[~0,~r_{\Sigma})$ and $\delta>0$, a small increment.\\

\section{\normalsize\bf{Space-time Matching }}

We shall now show the space-time matching of the collapsing star
with the exterior vacuum. First of all, we shall find the
matching conditions for a general exterior space-time and then
for Schwarzschild space-time as a particular case. For smooth
matching we shall use the Israel-Darmois [20] junction conditions
on the boundary $\sum$.\\

Let the interior of the star is denoted by $V^{-}$ and $V^{+}$ is
the exterior vacuum space-time region with $\sum$, the surface of
the star as the common boundary. So the interior space-time
($V^{-}$) is described by the metric given in equation (1) while
that of the exterior vacuum ($V^{+}$) is taken as

\begin{equation}
V^{+}:~~{ds_{+}}^{2}=-A^{2}(T,\Lambda)~dT^{2}+B^{2}(T,\Lambda)
~[d\Lambda^{2}+\Lambda^{2}d\Omega^{2}]
\end{equation}

The equation of the bounding surface $\sum$, considering it as an
embedding in the interior or exterior space-times are given by

\begin{equation}
\Sigma^{-}:~r=r_{\Sigma};~~~~~\Sigma^{+}:~\Lambda=\Lambda_{\Sigma}(T)
\end{equation}

with $r_{\Sigma}$, a constant. Then the metric in $\sum$ relative
to the coordinates of $V^{-}$ and $V^{+}$ are respectively

\begin{equation}
\Sigma^{-}:~{ds_{\Sigma^{-}}}^{2}=e^{-2\gamma}dt^{2}+R^{2}(t,r_{\Sigma})~d\Omega^{2}
\end{equation}

and

\begin{equation}
\Sigma^{+}:~{ds_{\Sigma^{+}}}^{2}=-A^{2}(T,\Lambda_{\Sigma})~dT^{2}+B^{2}
(T,\Lambda_{\Sigma})~[d{\Lambda_{\Sigma}}^{2}+{\Lambda_{\Sigma}}^{2}~d\Omega^{2}]
\end{equation}

According to Israel-Darmois, the matching of the interior and
exterior demands the continuity of the first and second
fundamental forms on the bounding hyper surface.(Note that, if
there are matters on the boundary then the second fundamental
form has a jump discontinuity over the boundary):\\

The continuity of the first fundamental form gives

\begin{equation}
i)~B\Lambda_{\Sigma}=R(t,r_{\Sigma})=R_{\Sigma}~~\text {(say)}~~
ii)~e^{\gamma}dt=\left[A^{2}-B^{2}\left(\frac{d\Lambda_{\Sigma}}{dT}\right)^{2}\right]^\frac{1}{2}
dT=d\tau
\end{equation}

For matching of the second fundamental form we need extrinsic
curvature which is given by

\begin{equation}
{K_{\mu\nu}}^{\pm}=-{n_{\sigma}}^{\pm}~{x^{\sigma}}_{,\zeta^{\mu},\zeta^{\nu}}
-{n_{\sigma}}^{\pm}~{\Gamma_{\beta\delta}}^{\sigma}{x^{\beta}}_{,\zeta^{\mu}}
{x^{\gamma}}_{,\zeta^{\nu}}
\end{equation}

where $\zeta^{\mu}$ are the intrinsic coordinates on $\sum$,
$x^{\sigma}$ are the coordinates of $V^{\pm}$, the (~,~) denotes
partial differentiation and the expression for the unit normal
$n_{\sigma}$ to the hypersurface ($f=$constant) is given by

\begin{equation}
n_{\sigma}=f_{,\sigma}/\left[g^{\mu\nu}f_{,\mu}f_{,\nu}\right]^{\frac{1}{2}}
\end{equation}

In the present matching model the explicit form of the unit
normal is given by

\begin{equation}
{n_{\sigma}}^{-}=\left(0,e^{\alpha(t,r_{\Sigma})},0,0\right)~~\text{and}
~~{n_{\sigma}}^{+}=\left(-\frac{d\Lambda_{\Sigma}}{dT},1,0,0\right)
\frac{AB}{\sqrt{A^{2}-B^{2}\left(\frac{d\Lambda_{\Sigma}}{dT}\right)^{2}}}
\end{equation}

Then the non-vanishing components of the extrinsic curvature in
$V^{-}$ and $V^{+}$ have the expressions

\begin{equation}
V^{-}:~{K_{\theta\theta}}^{-}=Sin^{-2}\theta{K_{\phi\phi}}^{-}
=(RR'e^{-\psi})_{\Sigma^{-}}
\end{equation}

\begin{eqnarray*}
V^{+}:~{K_{TT}}^{+}=\frac{AB}{\left(A^{2}-B^{2}\left(\frac{d\Lambda_{\Sigma}}{dT}
\right)^{2}\right)^{\frac{3}{2}}}\left\{\frac{B}{A^{2}}\frac{\partial
B}{\partial T}\left(\frac{d\Lambda_{\Sigma}}{dT}\right)^{3}
+\left(\frac{2}{A}\frac{\partial A}{\partial \Lambda}
-\frac{1}{B}\frac{\partial B}{\partial
\Lambda}\right)\left(\frac{d\Lambda_{\Sigma}}{dT}\right)^{2}\right.
\end{eqnarray*}
\begin{equation}
\left.+\left(\frac{1}{A}\frac{\partial A}{\partial
T}-\frac{2}{B}\frac{\partial B}{\partial
T}\right)\left(\frac{d\Lambda_{\Sigma}}{dT}\right)
-\frac{d^{2}\Lambda_{\Sigma}}{dT^{2}}-\frac{A}{B^{2}}
\frac{\partial A}{\partial \Lambda}\right\}_{\Sigma^{+}}
\end{equation}

and

\begin{equation}
{K_{\theta\theta}}^{+}=\frac{AB^{2}\Lambda_{\Sigma}}{\left(A^{2}
-B^{2}\left(\frac{d\Lambda_{\Sigma}}{dT}\right)^{2}\right)^{\frac{1}{2}}}
\left[\frac{\Lambda_{\Sigma}}{A^{2}}\left(\frac{\partial
B}{\partial T}\right)\frac{d\Lambda_{\Sigma}}{dT}+\frac{(B
+\frac{\partial B}{\partial
\Lambda}\Lambda_{\Sigma})}{B^{2}}\right]_{\Sigma^{+}}
\end{equation}

Hence matching of the second fundamental form demands

\begin{equation}
{K_{TT}}^{+}=0~~\text
{and}~~{K_{\theta\theta}}^{+}=(RR'e^{-\psi})_{\Sigma^{-}}
\end{equation}

In particular, if we take $V^{+}$ as the Schwarzschild metric,
namely,

\begin{equation}
V^{+}:~{ds_{+}}^{2}=-A(R)dT^{2}+\frac{1}{A(R)}dR^{2}+R^{2}d\Omega^{2}
\end{equation}

with $A(R)=1-\frac{2M}{R}$\\

then it can be seen easily by a straight forward calculation that

\begin{equation}
{K_{TT}}^{+}=0~~\text
{and}~~{K_{\theta\theta}}^{+}=\left(AR\frac{dT}{d\tau}\right)_{\Sigma^{+}}
\end{equation}

So the continuity of $K_{\theta\theta}$ over $\Sigma$ i.e,
${K_{\theta\theta}}^{+}-{K_{\theta\theta}}^{-}\mid_{\Sigma}=0$
simplifies to

\begin{equation}
2M=R_{\Sigma}\left[1-e^{-2\psi}R'^{2}+e^{-2\gamma}\dot{R}^{2}\right]_{\Sigma}
\end{equation}

which is the definition of the mass function given in equation
(11). Hence for a smooth matching of the interior (spherical
star) and exterior space-times (Schwarzschild), the interior mass
function at the surface must be equal to the Schwarzschild mass
of the exterior vacuum.\\

To study the formation of the horizon in this model we have from
equation (35)\\

~~~~~~~~~~~~~~~~~~~~~~~~~~~~~~~~~~~~~~~~~~~~~~~~~~~~~~~~$R=m(r)$\\

i.e, only one apparent horizon exists in this case.\\

In particular, if the exterior geometry is Schwarzschild, then
apparent horizon is the Schwarzschild horizon. Further, as
before, the time difference between the formation of apparent
horizon at comoving coordinate $r$ and the time of formation of
singularity of a shell at comoving coordinate $r$ is given by
$(r<<r_{\Sigma})$

\begin{equation}
t_{ah}(r)-t_{s}(r)=-{\int_{0}}^{v_{ah}(r)}\frac{\sqrt{v}dv}
{\sqrt{\left(1-\frac{2p_{t2}}{\rho_{0}}r^{2}v^{2}\right)\left(\frac{\rho_{0}}{3}
+b_{02}v-\frac{2p_{t2}}{\rho_{0}}b(r)v^{3}\right)}}
\end{equation}

Thus locally, if $t_{ah}(r)<t_{s}(r)$ then final state of
collapse will be a black hole while if $t_{ah}(r)\geq t_{s}(r)$
then final singularity will be naked. The above time difference
depends on the tangential pressure and total matter density. Thus
dark energy in the form of anisotropic fluid influences the
collapsing process and also the end state of collapse.\\

{\bf Acknowledgement}:\\

The author (S.C) is thankful to CSIR, Govt of India for providing
with a project no. 25(0141)/05/EMR-II on Gravitational Collapse.\\

{\bf References}:\\
\\
$[1]$ J.R.Oppenheimer and H.Snyder, {\it Phys.Rev.} {\bf 56}, 455
(1939). \\
$[2]$ P.S.Joshi and I.H.Dwivedi {\it Commun.Math.Phys.} {\bf 166},
117 (1994); {\it Class.Quantum.Grav.} {\bf 16}, 41 (1999).\\
$[3]$ K.Lake, {\it Phys.Rev.Lett.} {\bf 68}, 3129 (1992).\\
$[4]$ A.Ori and T.Piran, {\it Phys.Rev.Lett.} {\bf 59}, 2137
(1987).\\
$[5]$ T.Harada, {\it Phys.Rev.D} {\bf 58}, 104015 (1998).\\
$[6]$ U.Debnath, A.Banerjee and S.Chakraborty, {\it
Int.J.Mod.Phys.D} {\bf 12}, 1255 (2003).\\
$[7]$ P.S.Joshi, Global Aspects in Gravitation and Cosmology ({\it
Oxford Univ Press, Oxford,} 1993); See also the recent reviews:
P.S.Joshi, {\it Mod.Phys.Lett.A} {\bf 17}, 1067 (2002); A.Krolak
{\it Prog.Theor.Phys.Suppl.} {\bf 136}, 45 (1999); R.Penrose, in
Black Holes and Relativistic Stars, edited by R.Wald ({\it Univ.
of Chicago Press, Chicago}, 1998).\\
$[8]$ U.Debnath, Subenoy Chakraborty and N.Dadhich {\it
Int.J.Mod.Phys.D} {\bf 14}, 1761 (2005); Subenoy Chakraborty and
S.Chakraborty, {\it Int.J.Mod.Phys.D} (2005) (In Press); Subenoy
Chakraborty and S.Chakraborty, {\it Mod.Phys.Letts.A}
(2005) (In Press).\\
$[9]$ H.Muller zum Hagen, P.Yodzis and H.Seifert, {\it Commun.
Math.Phys.} {\bf 37}, 29 (1974).\\
$[10]$ L.Herrera and N.O.Santos, {\it Phys.Rept.} {\bf 286}, 53
(1997).\\
$[11]$ T.Harada, H.Iguchi and K.Nakao, {\it Prog.Theor.Phys.} {\bf
107}, 449 (2002).\\
$[12]$ Subenoy Chakraborty, S.Chakraborty and U.Debnath, {\it
Int.J.Mod.Phys.D} {\bf 14}, 1707 (2005).\\
$[13]$ R.Goswami and P.S.Joshi, {\it Class.Quantum.Grav.} {\bf
19}, 5229 (2002); T.Harada, K.Nakao and H.Iguchi, {\it
Class.Quantum.Grav.} {\bf 16}, 2785 (1999); S.M.C.V.Goncalves,
S.Jhingan and G.Magli, {\it Phys.Rev.D} {\bf 65}, 064011 (2002);
G.Magli, {\it Class.Quantum.Grav.} {\bf 14}, 1937 (1997); {\bf
15}, 3215 (1998).\\
$[14]$ S.S.Deshingkar, S.Jhingan, A.Chamorro and P.S.Joshi, {\it
Phys.Rev.D} {\bf 63}, 124005 (2001); M.Cissoko, J.C.Fabris,
J.Gariel, G.L.Denmat and N.O.Santos, {\it gr-qc}/9809057;
D.Markovic and S.L.Shapiro, {\it Phys.Rev.D} {\bf 61}, 084029
(2000); K.Lake, {\it Phys.Rev.D} {\bf 62}, 027301 (2000).\\
$[15]$ U.Debnath, S.Chakraborty and J.D.Barrow, {\it
Gen.Relt.Grav.} {\bf 36}, 231 (2004).\\
$[16]$ T.Arun Madhav, R.Goswami and P.S.Joshi, {\it Phys.Rev.D}
{\bf 72}, 084029 (2005).\\
$[17]$ P.M.Garnavich et al (Hi-Z Supernova Team Collaboration),
{\it Astrophys.J} {\bf 493}, L53 (1998); S.Perimutter et al
(Supernova Cosmology Project Collaboration), {\it Astrophys.J.}
{\bf 483}, 565 (1997); Nature (London), {\bf 391}, 51 (1998).\\
$[18]$ M.E.Cahill and G.C.McVittie, {\it J.Math.Phys.}(N.Y) {\bf
11}, 1382 (1970).\\
$[19]$ S.W.Hawking and G.F.R.Ellis, The Large Scale Structure of
Space-Time ({\it Cambridge Univ., Camb., UK}, 1973).\\
$[20]$ W.Israel, {\it Nuovo Cimento B} {\bf 44}, 1 (1966);
G.Darmois, Memorial des Sciences Mathematiques (Gauthier-Villars,
Paris, 1927), {\it Fasc.} 25; W.B.Bonnor and P.A.Vickers, {\it
Gen.Relt.Grav.} {\bf 13}, 29 (1981).\\

\end{document}